\documentclass[aps,prl,twocolumn]{revtex4}
\usepackage{graphicx,color}
\usepackage{bm}
\usepackage{dsfont,amsmath,amssymb}
\usepackage{amsthm}

\newcommand{\Tr}{\mathrm{Tr}}

\renewcommand{\SS}{\mathrm{S}}
\newcommand{\RR}{\mathrm{R}}
\newcommand{\SR}{\mathrm{SR}}

\begin{document}
\title{Strong Coupling Thermodynamics of Open Quantum Systems}
\author{\'{A}ngel Rivas}
\affiliation{Departamento de F\'isica Te\'orica, Facultad de Ciencias F\'isicas,
Universidad Complutense, 28040 Madrid, Spain.\\
CCS -Center for Computational Simulation, Campus de Montegancedo UPM, 28660 Boadilla del Monte, Madrid, Spain.}

\date{\today}

\begin{abstract}
A general thermodynamic framework is presented for open quantum systems in  fixed contact with a thermal reservoir. The first and second law are obtained for arbitrary system-reservoir coupling strengths, and including both factorized and correlated initial conditions. The thermodynamic properties are adapted to the generally strong coupling regime, approaching the ones defined for equilibrium, and their standard weak-coupling counterparts as appropriate limits. Moreover, they can be inferred from measurements involving only system observables. Finally, a thermodynamic signature of non-Markovianity is formulated in the form of a negative entropy production rate.
\end{abstract}

\maketitle

\paragraph{Introduction.---}
The statement of thermodynamics laws at the quantum level is an open and fundamental task. However, given its practical implications in areas such as quantum transport \cite{QTrans1,QTrans2,QTrans3}, quantum information \cite{QI1,QI2,QI3}, or AMO physics \cite{AMO1,AMO2,AMO3,AMO4,AMO5,AMO6,AMO7,AMO8,AMO9,AMO10}, the motivation is far from being only fundamental. For these reasons, the field of quantum thermodynamics has emerged lately attracting an extensive attention \cite{QT1,QT2}. 

A particularly interesting problem concerns the formulation of a universally valid nonequilibrium thermodynamic framework for open quantum systems in contact with thermal reservoirs \cite{Kosloff1}. By ``reservoir'' we understand a quantum system with an infinitely large, continuous, number of degrees of freedom, and a ``thermal reservoir'' is the one which initially remains in its canonical Gibbs state $\rho_\RR(0)=\rho_{\RR,\beta}=\exp(-\beta H_\RR)/Z_\RR$. Here, $\beta=1/(k_B T)$ is the inverse temperature (with $T$ temperature and $k_B$ Boltzmann constant), $H_\RR$ is the reservoir Hamiltonian and $Z_\RR=\Tr[\exp(-\beta H_\RR)]$ is its partition function. Strictly speaking, since $Z_\RR$ becomes infinity for an infinitely large, continuous, system, the density matrix $\rho_{\RR,\beta}$ is ill-defined in such a case. The rigorous definition of this state is given as a functional in the algebraic formulation of quantum mechanics \cite{Bratteli}. However, we shall write $\rho_{\RR,\beta}$ with a formal meaning. If the interaction Hamiltonian between system and reservoir is denoted by $V$, the total Hamiltonian reads
\begin{equation}\label{H}
H(t)=H_\SS(t)+H_\RR+V,
\end{equation}
where $H_\SS(t)$ is a generally time-dependent system Hamiltonian (unless otherwise stated we shall adopt the Schr\"odinger picture throughout the text). Considering the system and thermal reservoir initially in the product state
\begin{equation}\label{InitialState}
\rho_\SR(0)=\rho_\SS(0)\otimes\rho_{\RR,\beta},
\end{equation}
after some time interval $t$, the state changes to
\begin{equation}
\rho_\SR(t)=U(t,0)\rho_\SS(0)\otimes\rho_{\RR,\beta} U^\dagger(t,0),
\end{equation}
with the evolution family $U(t,0):=\mathcal{T}\exp\big[-\frac{i}{\hbar} \int_0^t H(s) ds\big]$, where $\mathcal{T}$ is the time-ordering operator. This dynamics induces a time-evolution in the open system $\SS$ given by a dynamical map $\Lambda_t$, i.e. a family of completely positive and trace-preserving (CPTP) maps \cite{AlickiBook,BrPe02,Libro}, $\rho_\SS(0)\to \rho_\SS(t)=\Lambda_t\rho_\SS(0):=\Tr_\RR[\rho_{\SR}(t)]$. We shall address the derivation of the thermodynamics laws for the open quantum system $\SS$ in this situation.

\paragraph{Weak coupling considerations.---}
The first step is the identification of system thermodynamic variables. Since the global system composed by open system and reservoir is isolated, any energy change (which only occurs for time-dependent Hamiltonians) must be identified with work $W$. Thus, the power is given by
\begin{equation}\label{Workdrf}
\dot{W}(t):=\frac{d\langle H(t)\rangle}{dt}=\Tr[\dot{H}_\SS(t)\rho_\SR(t)]=\Tr[\dot{H}_\SS(t)\rho_\SS(t)],
\end{equation}
where, in the second equality, we have used the von Neumann equation $\dot{\rho}_\SR(t)=-\tfrac{i}{\hbar}[H(t),\rho_\SR]$, and adopted the overdot notation for time-derivatives. This work is assumed to be performed by/applied to the system as only depends on system  variables. 

Internal energy and heat are magnitudes more difficult to be properly defined. However, this task can be successfully accomplished in the limit of small interaction $V$. In such a case, the expectation value of the total Hamiltonian becomes $\langle H(t)\rangle \simeq \langle H_\SS(t)\rangle+\langle H_\RR\rangle$, and so $\langle H_\SS(t) \rangle$ can be unequivocally identified with the system internal energy $E_{\rm U}$ at time $t$ \cite{Footnote1}. Then, taking time-derivative one obtains the first law in the form of 
\begin{equation}\label{1LawWeak}
\frac{dE_{\rm U}^{\rm (w)}(t)}{dt}=\frac{d\langle H_\SS(t)\rangle}{dt} =\dot{W}(t)+\dot{Q}^{\rm(w)}(t),
\end{equation}
with $\dot{Q}^{\rm (w)}(t):=\Tr[H_\SS(t) \dot{\rho}_\SS(t)]$ the heat flow in the weak coupling approximation.

The second law can also be obtained in the weak coupling limit. For a slow time-varying $H_\SS(t)$ compared to the relaxation time of the reservoir \cite{Alicki79}, the dynamical map can be rigorously approximated by
$
\Lambda_t=\mathcal{T}\exp\big[\int_0^{t}\mathcal{L}_{\rm D}(s)ds\big]
$
where $\mathcal{L}_{\rm D}(t)$ is the time-dependent ``Davies generator'' \cite{DaviesSpohn} with the (time-dependent) GKLS form \cite{GKLS1,GKLS2} so that $\Lambda_t$ is CPTP. Thanks to this, and applying a series of results \cite{Kosloff1,SpohnEntropy} based on the monotonicity of the quantum relative entropy $D(\rho_1\Vert \rho_2):=\Tr(\rho_1\log\rho_1)-\Tr(\rho_1\log\rho_2)$ under a CPTP map $\Lambda$ \cite{MonoCP1,MonoCP2},
\begin{equation}\label{Monotonicity}
D[\Lambda(\rho_1)\Vert\Lambda(\rho_2)]\leq D(\rho_1\Vert \rho_2),
\end{equation}
it is possible to obtain the second law in the differential form
\begin{equation}\label{Srate}
\frac{dS_{\rm vN}(t)}{dt}-\beta \dot{Q}^{\rm (w)}(t)\geq 0,
\end{equation}
where $S_{\rm vN}(t):=-k_B \Tr[\rho_\SS(t)\log\rho_\SS(t)]$ is the (thermodynamic) von Neumann entropy. This can be extended to arbitrarily fast periodic drivings \cite{Kosloff1,AlickiPeriodic1,AlickiPeriodic2}. Recently, other drivings have also been analyzed \cite{Kosloff2}. 

Out of the weak coupling regime, several attempts have been performed to formulate a thermodynamic framework, e.g. \cite{EspositoEntropy,Modi,StrasbergNJP,Alipour,Tanimura,StrasbergPRX,Ghosh,Eisert1,Hsiang,StrasbergPRE,RefinedEntropy,StrasbergArxiv}. A possible approach \cite{EspositoEntropy,Tanimura,StrasbergPRX} defines $\dot{Q}^{\rm (e)}(t):=-\Tr[H_\RR(t)\dot{\rho}_\RR(t)]$ and $E_{\rm U}^{\rm (e)}(t):=\Tr\{[H_\SS(t)+V]\rho_\SR(t)\}$ as heat and system internal energy, respectively. Here, the superscript ``e'' stands for ``external'' as, in this approach, those variables are actually external properties, defined in terms of reservoir mean values. This is undesirable from the open system theory, and implies the experimental difficulty of controlling the state of the reservoir in order to make thermodynamic measurements. Nevertheless, it is possible to obtain the first and the second law in the integrated form $\Delta S_{\rm vN}(t)-\beta Q^{(\rm e)}(t)\geq0$ for a ``finite size'' reservoir \cite{EspositoEntropy,Tanimura,StrasbergPRX}. Although they are also expected to hold for true (infinite) reservoirs as an appropriate limit, one should be careful at this point in the continuous limit because quantities such as the reservoir von Neumann entropy are ill-defined. On the other hand, as we shall see in a moment, these definitions do not fit with the expected situation once the system reaches thermal equilibrium. In order to overcome these difficulties an alternative approach is needed.

\paragraph{Equilibrium considerations.---}
Let us assume for a moment that $H_\SS$ is time-independent in \eqref{H}. There is a vast literature showing that system and reservoir thermalize after some transient time interval \cite{Bach,Jaksic1,Jaksic2,Merkli0, Merkli1,Merkli2,Merkli3,Merkli4, Linden,Reimann,Short,Eisert2,HuConvergence,StrasbergNJP,Nazir}. In particular, under certain regularity conditions on reservoirs and couplings, it can been rigorously proven \cite{Bach,Jaksic1,Jaksic2,Merkli0, Merkli1,Merkli2,Merkli3,Merkli4} that
\begin{equation}\label{ConvH}
\rho_\SR(t)=e^{-i H t/\hbar}\rho_\SS(0)\otimes\rho_{\RR,\beta} e^{i H t/\hbar}\xrightarrow{t\to \infty}\frac{e^{-\beta H}}{Z_\SR}.
\end{equation}  
This convergence requires $\RR$ to be an infinitely large continuum, and this implies $Z_\SR=\Tr[\exp(-\beta H)]$ to be singular, as commented. Thus, the limit in \eqref{ConvH} must be understood in functional sense \cite{Footnote2}. 
Thermalization suggests that, any suitable choice of nonequilibrium system internal energy must fit the system internal energy obtained from the global canonical state $\rho_{\SR,\beta}=Z_\SR^{-1}\exp(-\beta H)$ once equilibration is reached (perhaps asymptotically). That equilibrium thermal internal energy has been studied both in the classical \cite{Kirkwood,Gelin,Jarzynski,Seifert,Anders,StragsberClass} and in the quantum \cite{Gelin,Hanggi, Hu,StrasbergPRE} realm. The method is based on the definition of the ``Hamiltonian of mean force'' $H_{\SS}^*$ by the equation
\begin{equation}\label{Hmean}
H_\SS^*:=-\beta^{-1}\log\Big[\big(\tfrac{Z_\SR}{Z_\RR}\big)\Tr_\RR(\rho_{\SR,\beta})\Big],
\end{equation}
so that the reduced system state at thermal equilibrium is given formally by a Gibbs state for $H^*_{\SS}$,
\begin{equation}\label{rhoeq}
\rho_{\SS,\mathrm{eq}}:=\Tr_\RR(\rho_{\SR,\beta})=\frac{e^{-\beta H_\SS^*}}{Z^*_\SS},
\end{equation}
with $Z^*_\SS=Z_\SR/Z_\RR$. Now, one requires the fulfillment of standard equilibrium relations such us $F=-\beta^{-1}\log Z_\SS$, $E_{\rm U}=F+\partial_\beta F$, $S=\beta^2\partial_\beta F$, and $F=E_{\rm U}-TS$, for $Z^*_\SS$ playing the role of $Z_\SS=\Tr[\exp(-\beta H_\SS)]$ ($Z^*_\SS$ approaches  $Z_\SS$ for vanishing coupling). Since the Hamiltonian of mean force \eqref{Hmean} is a function of $\beta$, $H_{\SS}^*(\beta)$, this leads to the following redefinitions of internal and free energy and thermodynamic entropy at equilibrium (units of $k_B=1$):
\begin{align}
& E_{\rm U}^*:=\Tr\{\rho_{\SS,\mathrm{eq}} [H_\SS^*(\beta)+\beta \partial_\beta H_\SS^*(\beta)]\}, \label{E*}\\
& F^*:=\Tr\{\rho_{\SS,\mathrm{eq}} [H_\SS^*(\beta)+\beta^{-1} \log \rho_{\SS,\mathrm{eq}}]\},\label{F*}\\
& S^*:=\Tr\{\rho_{\SS,\mathrm{eq}} [-\log\rho_{\SS,\mathrm{eq}}+\beta^2\partial_\beta H_\SS^*(\beta)]\}.\label{S*}
\end{align}
A possible generalization of Eqs. \eqref{E*}-\eqref{S*} for nonequilibrium is given by the straightforward substitution $\rho_{\SS,\mathrm{eq}}\to \rho_{\SS}(t)$ \cite{Seifert,StrasbergPRE,StragsberClass}. This choice satisfies thermodynamic laws for some restricted class of initial states \cite{StrasbergPRE}, but fails for the general initial condition \eqref{InitialState} \cite{SM}. Therefore, we shall take a different route.

\paragraph{Time-independent system Hamiltonians.---} 
Let us consider first a time-independent $H_\SS$. Typically, $\Tr_\RR(V\rho_{\RR,\beta})=0$ (otherwise this can always be achieved by a convenient redefinition of system and interaction Hamiltonians, see e.g. \cite{Libro}), and so under the initial condition \eqref{InitialState} we have 
\begin{equation}\label{HSep}
\langle H(0)\rangle=\Tr[H \rho_\SS(0)\otimes\rho_{\RR,\beta}]=\Tr[H_\SS \rho_\SS(0)]+\Tr[H_\RR \rho_{\RR,\beta}].
\end{equation}
The first term of the right hand side of \eqref{HSep} can be unambiguously identified with the system internal energy at $t=0$. In addition, because of \eqref{ConvH}, the internal energy must become equivalent to \eqref{E*} for asymptotic times. Strictly speaking, since only relative differences appear in the thermodynamic laws, this equivalence must hold up to some time-independent additive constant (which observes $E_{\rm U}=F+\partial_\beta F$, $S=\beta^2\partial_\beta F$, $F=E_{\rm U}-TS$ at thermal equilibrium). The third property we desire for the choice of internal energy is that it should be given just in terms of the reduced system dynamics. In similar spirit to \cite{RefinedEntropy}, all these properties are satisfied by defining
\begin{equation}\label{Hcir1}
H^\circledast_\SS(t,\beta):=-\beta^{-1}\log[\Lambda_t e^{-\beta H_\SS}],
\end{equation}
such that $\Lambda_t e^{-\beta H_\SS}=e^{-\beta H_\SS^\circledast(t,\beta)}$ and, in parallelism  with \eqref{E*}-\eqref{S*},
\begin{align}
& E_{\rm U}(t):=\Tr\{\rho_\SS(t) [H_\SS^\circledast(t,\beta)+\beta \partial_\beta H_\SS^\circledast(t,\beta)]\}, \label{Eexact}\\
& F(t):=\Tr\{\rho_\SS(t) [H_\SS^\circledast(t,\beta)+\beta^{-1} \log \rho_\SS(t)]\}, \label{Fexact}\\
& S(t):=\Tr\{\rho_\SS(t) [-\log\rho_\SS(t)+\beta^2\partial_\beta H_\SS^\circledast(t,\beta)]\}. \label{Sexact}
\end{align}
One should note that
\begin{align}
&H_\SS^\circledast(0,\beta)=H_\SS,\label{Hcirc0}\\
&H_\SS^\circledast(\infty,\beta)=H_\SS^*+\beta^{-1}\log[Z_{\SR}/(Z_{\SS}Z_{\RR})], \label{HcircInf}
\end{align}
where the last additive constant is due to the fact that $\Lambda_t$ is trace-preserving \cite{Footnote3}. These relations ensure the correct initial and long time limits of $E_{\rm U}(t)$. Moreover, for small coupling $V$, the dynamics is given by the Davies semigroup \cite{AlickiBook,BrPe02,Libro} which has $\exp(-\beta H_\SS)$ as a fixed point, and so $E_{\rm U}(t)\to E_{\rm U}^{\rm (w)}(t)$ and the thermodynamic entropy $S(t)\to-\Tr[\rho_\SS(t)\log\rho_\SS(t)]$ approaches the usual von Neumann expression.  

By construction, the definitions \eqref{Eexact}-\eqref{Sexact} satisfy asymptotically, at thermal equilibrium, the `standard' relations $F(\infty)=-\beta^{-1}\log Z_\SS^\circledast(\infty)$, $E_{\rm U}(\infty)=F(\infty)+\partial_\beta F(\infty)$, and $S(\infty)=\beta^2\partial_\beta F(\infty)$. In addition, since $\Lambda_t$ is trace preserving, $Z_{\SS}^\circledast(\infty):=\Tr\{\exp[-\beta H_\SS^\circledast(\infty,\beta)]\}=\Tr[\exp(-\beta H_\SS)]=Z_\SS$. This implies that the thermodynamic variables at equilibrium take the same value regardless of the strength of the coupling $V$ \cite{Footnote3bis}, so they can be obtained by e.g. their weak coupling expressions. Namely, for the internal energy we have $E_{\rm U}(\infty)=-\partial_\beta\log Z_\SS^\circledast(\infty)=-\partial_\beta\log Z_\SS=E^{\rm (w)}_{\rm U}(\infty)=\Tr(\rho_{\SS,\beta}H_\SS)$. Similarly, $S(\infty)=-\Tr(\rho_{\SS,\beta}\log\rho_{\SS,\beta})$ and so the entropy \eqref{Sexact} at equilibrium approaches 0 for vanishing temperature (in absence of degeneracy), as expected for a `thermodynamic' entropy. This behavior has also been found for $S^\ast$ \cite{Hu}, but it is not fulfilled for the von Neumann entropy as entanglement may preclude the reduced state to be pure for non-vanishing $V$.

On the other hand, since for a time-independent Hamiltonian $H_\SS$ there is no work, the \emph{first law} defines heat as
\begin{align}\label{1lawHInd}
\dot{Q}(t)=\frac{dE_{\rm U}(t)}{dt}\Rightarrow Q(t)&=E_{\rm U}(t)-E_{\rm U}(0)\nonumber \\
&=E_{\rm U}(t)-\langle H_\SS(0)\rangle.
\end{align}
In regard to the \emph{second law}, it can be derived in the integrated form. From Eq. \eqref{Monotonicity},
\begin{equation}\label{2lawHInd_A1}
D\Big\{\Lambda_{t}[\rho_{\SS}(0)]\Big\Vert\Lambda_{t}\big(\rho_{\SS,\beta}\big)\Big\}\leq D\Big[\rho_{\SS}(0)\Big\Vert \rho_{\SS,\beta}\Big],
\end{equation}
which can be straightforwardly recast in the form
\begin{multline}\label{2lawHInd_A2}
-\Tr[\rho_\SS(t)\log\rho_\SS(t)]-S(0)\\
-\beta \{\Tr[\rho_\SS(t)H_\SS^\circledast(t,\beta)]-\langle H_\SS(0)\rangle \}\geq 0.
\end{multline}
By adding and subtracting $\beta^2\Tr[\rho_\SS(t)\partial_\beta H_\SS^\circledast(t,\beta)]$, and using \eqref{1lawHInd}, we finally obtain
\begin{equation}\label{2lawHInd}
\Delta S(t)-\beta Q(t)\geq 0.
\end{equation}

One may notice that this equation for the entropy production reaches the zero value if the system is initially in the Gibbs state $\rho_\SS(0)=\rho_{\SS,\beta}$, as in that case the equality in \eqref{2lawHInd_A1} is trivially obtained. Actually, in such a situation, no thermodynamic magnitude in \eqref{Eexact}-\eqref{Sexact} changes on time. This might seem surprising but it can be understood because, formally, 
\begin{equation}
U(t,0)\rho_{\SS,\beta}\otimes\rho_{\RR,\beta} U^\dagger(t,0)=\frac{e^{-\beta U(t,0)(H_\SS+H_\RR)U^\dagger(t,0)}}{Z_\SS Z_\RR}.
\end{equation}
Since at $t=0$ \eqref{HSep} holds, system and reservoir starts effectively and remains canonical throughout the process, at instantaneous `thermal equilibrium' in the Gibbs state of the `Hamiltonian' $U(t,0)(H_\SS+H_\RR)U^\dagger(t,0)$. Given that, by hypothesis, there is no applied work, the internal energy and the rest of the thermodynamic properties remain constant. This `thermodynamic' reversibility should be considered, in the strong coupling, as a different concept from `informational' reversibility. The density matrix indeed changes on time despite the thermodynamic variables and so the thermodynamic `state' remain constant. However, this does not contradict the standard reversibility notion in weak-coupling thermodynamics. That is a particular case of this more general formalism where both informational variables (e.g. the von Neumann entropy) and thermodynamic variables \eqref{Eexact}-\eqref{Sexact} coincide.

\begin{figure}[t]
	\includegraphics[width=\columnwidth]{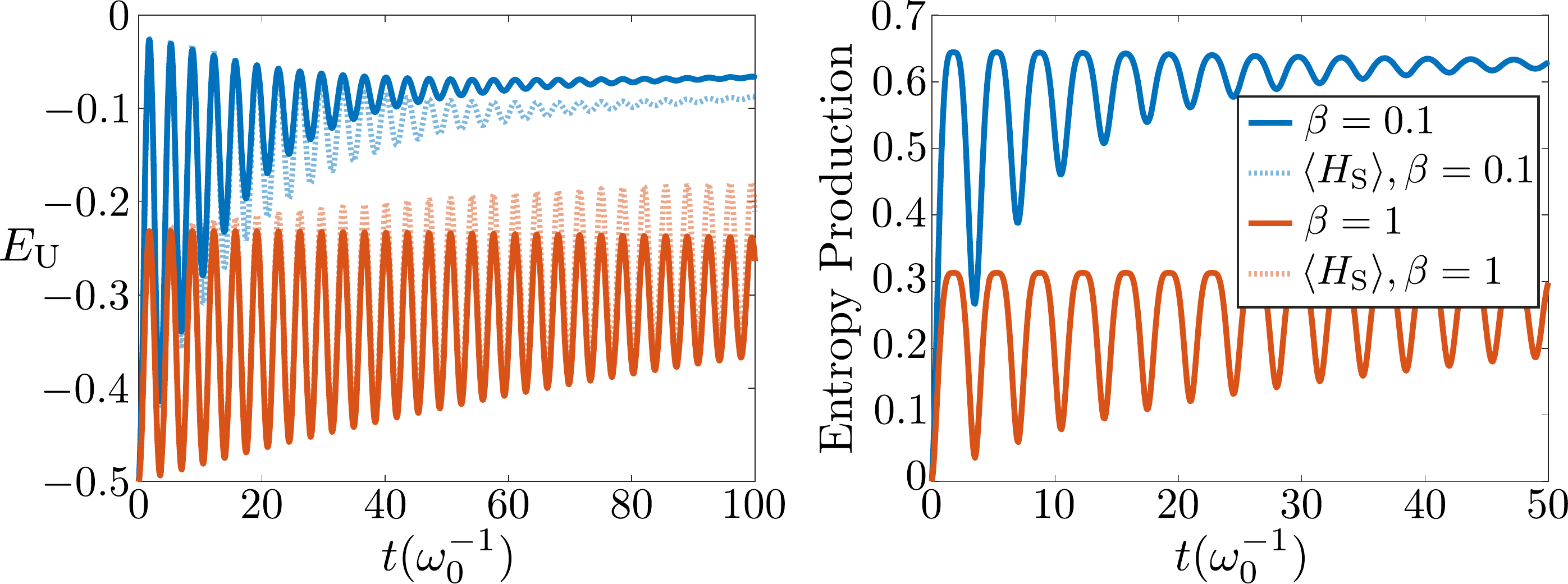}
	\caption{Internal energy (left) and entropy production (right) of a qubit interacting with a composite spin-boson reservoir. The qubit is initially in its groud state. The internal energy $E_{\rm U}$ turns out to be a modest correction to the expectation value of the system Hamiltonian $\langle H_\SS\rangle$ (lighter color, dotted line). The nonmonotonic entropy production \eqref{2lawHInd} shows the non-Markovian character of the dynamics, see Eq. \eqref{entropyprod_rate}. We have taken $\omega_0=\omega_{1}$, $\kappa=0.9\omega_0$, and $10^{-3}\omega_0$ for the spin-boson decay rates \cite{SM}.}
	\label{Fig:1}
\end{figure}

\paragraph{Time-dependent system Hamiltonians.---} 
For a general time-dependent $H_\SS(t)$, we still demand the same initial condition for the internal energy $E_{\rm U}(0)=\langle H_\SS(0)\rangle=\Tr[H_\SS(0)\rho_\SS(0)]$ as before, and the recovery of the previous results if $H_\SS(t)$ becomes time-independent. Moreover, we retain the requirement for a formulation just in terms of system observables. This can be done by keeping the definitions \eqref{Eexact}-\eqref{Sexact} and redefining the operator $H_\SS^\circledast(t,\beta)$ by 
\begin{equation}\label{Hcirct}
H^\circledast_\SS(t,\beta):=-\beta^{-1}\log\left[\Lambda_{t}\left\{e^{-\beta H_{\SS}(0)-\beta\int_0^t \Lambda_{s}^\star[\dot{H}_{\SS}(s)]ds}\right\}\right]
\end{equation}
where $\Lambda^{\star}$ denotes the Heisenberg adjoint of $\Lambda$, $\Tr[\Lambda(A)B]=\Tr[A\Lambda^\star(B)]$ \cite{RefinedEntropy}. Note that $H^\circledast_\SS(0,\beta)=H_\SS(0)$ and for time-independent $H_\SS$ the equation \eqref{Hcir1} is recovered, as required.

The \emph{first law} in this case reads 
\begin{align}\label{1lawH}
\frac{d E_{\rm U}(t)}{dt}=\dot{Q}(t)+\dot{W}(t),
\end{align}
and that is used to define heat,
\begin{align}\label{Heat}
Q(t):=E_{\rm U}(t)-\langle H_\SS(0)\rangle-\int_0^t\Tr[\rho_{\SS}(s)\dot{H}_{\SS}(s)]ds.
\end{align}
Note that, if we consider a quasistatic and small coupling regime where {\small $\Lambda_t\Big(\exp\big\{-\beta H_{\SS}(0)-\beta\int_0^t \Lambda_{s}^\star[\dot{H}_{\SS}(s)\big]ds\big\}\Big)\approx \exp[-\beta H_\SS(t)]$} such that $H^\circledast_\SS(t,\beta)\approx H_\SS(t)$, the weak-coupling first law \eqref{1LawWeak} is obtained. This reinforces the definition \eqref{Hcirct}.

In order to derive the second law, we define the auxiliary object
\begin{equation}
\Omega(t,r):=-\beta^{-1} \log \left[\Lambda_{t}\left\{e^{-\beta H_{\SS}(0)-\beta\int_0^r \Lambda_{s}^\star[\dot{H}_{\SS}(s)]ds}\right\}\right],
\end{equation}
which satisfies $\Omega(t,t)=H_{\SS}^\circledast(t,\beta)$. A straightforward computation in \eqref{Heat} gives 
\begin{equation}\label{RefQaux}
Q(t)=\Tr\{\rho_{\SS}(t)[\Omega(t,t)+\beta\partial_\beta H_\SS^\circledast(t,\beta)]\}-\Tr[\rho_{\SS}(0)\Omega(0,t)].
\end{equation}
For the state
\begin{equation}
\rho_{0}(\beta,r):=Z^{-1}_\SS(r)e^{-\beta H_{\SS}(0)-\beta\int_0^r \Lambda_{s}^\star[\dot{H}_{\SS}(s)]ds},
\end{equation}
with {\small $Z_\SS(r)=\Tr\Big(\exp\big\{-\beta H_{\SS}(0)-\beta\int_0^r \Lambda_{s}^\star[\dot{H}_{\SS}(s)\big]ds\big\}\Big)$}, monotonicity of the relative entropy \eqref{Monotonicity} yields
\begin{equation}
D\Big\{\Lambda_{t}[\rho_{\SS}(0)]\Big\Vert\Lambda_{t}[\rho_{0}(\beta,r)]\Big\}\leq D\Big[\rho_{\SS}(0)\Big\Vert \rho_{0}(\beta,r)\Big],
\end{equation}
which can be recast in the form
\begin{multline}
-\Tr[\rho_\SS(t)\log\rho_\SS(t)]-S(0)\\
-\beta\left\{\Tr[\rho_{\SS}(t)\Omega(t,r)]-\Tr[\rho_{\SS}(0)\Omega(0,r)]\right\}\geq 0.
\end{multline}
Since this is fulfilled for all $r$, and particularly for $r=t$, again by adding and subtracting $\beta^2\Tr[\rho_\SS(t)\partial_\beta H_\SS^\circledast(t,\beta)]$ and using \eqref{RefQaux}, we obtain the \emph{second law}
\begin{equation}\label{2lawH}
\Delta S(t)-\beta Q(t)\geq0.
\end{equation}
This completes the thermodynamic formulation for general open quantum systems in contact with a thermal reservoir.

As a simple example of this approach we may consider a qubit with Hamiltonian $H_{\rm S}=\tfrac{\omega_0}{2}\sigma_{\SS}^z$, in dipolar contact $V=\kappa (\sigma^+_{\SS}\sigma^-_{\rm R}+\sigma^-_{\SS}\sigma^+_{\RR})$ with a composite spin-boson reservoir \cite{Neil} with Hamiltonian $H_{\RR}=H_{\rm spin}+H_{\rm boson}+V_{\rm spin-boson}$. Here, $H_{\rm spin}=\tfrac{\omega_{1}}{2}\sigma^z_{\RR}$, $H_{\rm boson}=\int dk \omega(k) a^\dagger(k)a(k)$, $V_{\rm spin-boson}=\alpha \int dk g(k)\sigma^x_{\RR}[a(k)+a^\dagger(k)]$, $a(k)$ denotes bosonic anihilation operators, and $\sigma^{z,\pm}_{\SS}$, $\sigma^{z,x,\pm}_{\RR}$ stand for Pauli matrices for system and spin part of the reservoir, respectively. Note that the bosonic part of the reservoir is not directly coupled to the system qubit. The coupling between system and reservoir is mediated by $\kappa$, whereas $\alpha$ is supposed small enough such that a weak coupling treatment of the spin-boson degrees of freedom is justified \cite{SM}.  Figure \ref{Fig:1} shows the comparison between the internal energy $E_{\rm U}$, Eq. \eqref{Eexact}, in the strong coupling and the mean value of the the system Hamiltonian $\langle H_\SS\rangle$ for several temperatures. One can notice that, for this model, the latter turns out to be a modest correction to $\langle H_\SS\rangle$ sharing a very similar time-dependency.

\paragraph{Initially correlated states.---}
It is worth to examine whether the previous approach can be extended to different initial system-reservoir states. We do not expect that for any initial state, but for those sufficiently close to the thermodynamic paradigm of a system coupled to a thermal reservoir. Namely, we should consider just those initial system-reservoir states where the reservoir can be well-described via the temperature parameter $\beta$. This condition can be rigorously formulated in the framework of operator algebras \cite{Bach,Merkli1,Merkli2,Merkli3}, but, for our purposes, there are another two natural classes of states which can be considered in addition to \eqref{InitialState}. They correspond to the displacement from the global equilibrium $\rho_{\SR,\beta}$ either by system driving $H_{\SS}(t)$ or by system quantum measurements, respectively \cite{StrasbergPRE}.

For the first case $\rho_{\SR}(0)=\rho_{\SR,\beta}$, and we can assume, formally, a former product `initial' condition $\rho_{\SS,\beta}\otimes\rho_{\RR,\beta}$ at $t_0=-\infty$ and $\dot{H}_\SS(t)=\dot{H}_{\SS}(t)\theta(t)$, with $\theta(t)$ the step function. Following the same steps as before we conclude
\begin{equation}
S(t)-S(-\infty)-\beta Q_{-\infty}(t)\geq 0,
\end{equation}
where $Q_{-\infty}(t):=\int_{-\infty}^t\dot{Q}(s)ds$ is the heat in the interval $(-\infty,t)$. Then, by splitting this integral into positive and negative time values, and adding and subtracting $S(0)$, which in this case is the entropy \eqref{Sexact} of the reduced equilibrium state \eqref{rhoeq}, we have
\begin{multline}
S(t)-S(0)+S(0)-S(-\infty)\\
-\beta \left[\int_{0}^{t}\dot{Q}(s)ds +\int_{-\infty}^{0}\dot{Q}(s)ds\right]\geq0.
\end{multline} 
Since we have taken $\dot{H}_{\SS}(t)=0$ for $t<0$, and the entropy production for the canonical system state $\rho_{\SS,\beta}$ reaches the zero value for time-independent system Hamiltonians, $S(0)-S(-\infty)-\beta \int_{-\infty}^{0}\dot{Q}(s)ds=0$. Hence, we obtain the desired result \eqref{2lawH} for $\rho_{\SR}(0)=\rho_{\SR,\beta}$.

For the second case, the joint initial state (after a generally nonselective, projective measurement) is written as
\begin{equation}\label{0discord}
\rho_{\SR}(0)=\sum_{k}p_k \Pi_k \otimes \rho_{\RR|k}
\end{equation}
with $\Pi_k=|k\rangle\langle k|$ a complete set of orthonormal projectors and
\begin{align}
p_k=\Tr\big(\Pi_k\otimes \mathbb{I}\rho_{\SR,\beta}\big), \quad \rho_{\RR|k}=\frac{\Tr_\SS\big(\Pi_k\otimes \mathbb{I}\rho_{\SR,\beta}\big)}{p_k}.
\end{align}
For this kind of states it is possible to write the reduced system dynamics as $\tilde{\Lambda}_t\rho_{\SS}(0)=\rho_\SS(t)$ for $\rho_\SS(0)=\sum_k p_k \Pi_k$, with $\tilde{\Lambda}_t$ a CPTP map \cite{Cesar,Footnote4}.

On the other hand, $\rho_{\SR,\beta}$ remains static before the measurement, with system internal energy \eqref{Eexact} given by
\begin{equation}
E_{\rm U}({\rm eq}):=\Tr\{\rho_{\SS,\rm eq} [H_\SS^\circledast({\rm eq},\beta)+\beta \partial_\beta H_\SS^\circledast({\rm eq},\beta)]\},
\end{equation}
where $H_\SS^\circledast({\rm eq},\beta)=H_\SS^*+\beta^{-1}\log[Z_{\SR}/(Z_{\SS}Z_{\RR})]$ according to \eqref{HcircInf}. Therefore, it seems reasonable to take the internal energy after the measurement as
\begin{equation}
E_{\rm U}(0)=\Tr\{\rho_\SS(0) [H_\SS^\circledast({\rm eq},\beta)+\beta \partial_\beta H_\SS^\circledast({\rm eq},\beta)]\},
\end{equation}
with $\rho_\SS(0)=\sum_k p_k \Pi_k$. A finer choice could be possible with a microscopic model for the measurement interaction where the measurement change was not `instantaneous'. Then by redefining $H_\SS^{\circledast}(t,\beta)$ as in \eqref{Hcirct} with $\tilde{\Lambda}_t$ and $H_\SS^\circledast({\rm eq},\beta)$ in the roles of $\Lambda_t$ and $H_\SS(0)$, respectively, the derivation of \eqref{2lawH} follows from the same argument as in previous sections.

\paragraph{Non-Markovianity.---}
Finally, we show that, within this approach, it is possible to establish a thermodynamic signature of non-Markovianity (see also \cite{Misra,Alipour2,StrasbergPRE,StrasbergArxiv}). Suppose the dynamical map $\Lambda_t$ to be CP divisible \cite{revMarko1,revMarko2,revMarko3} (actually P-divisible is enough \cite{Reeb}); namely, it can be decomposed as $\Lambda_t=\Lambda_{t,s}\Lambda_{s}$ for any pair $t>s$ with $\Lambda_{t,s}$ CPTP. Then, monotonicity of the relative entropy \eqref{Monotonicity} implies 
\begin{multline}
S\Big\{\Lambda_{t+\epsilon}[\rho_{\SS}(0)]\Big\Vert\Lambda_{t+\epsilon}[\rho_{\mathrm{S},\beta}]\Big\}\\
\leq S\Big\{\Lambda_{t}[\rho_{\SS}(0)]\Big\Vert\Lambda_{t}[\rho_{\mathrm{S},\beta}]\Big\},
\end{multline}
for $\epsilon>0$. From here, following similar steps as for \eqref{2lawHInd} and dividing by $\epsilon$ in the limit $\epsilon\to 0$, we obtain a positive entropy production rate
\begin{equation}\label{entropyprod_rate}
\frac{dS(t)}{dt}-\beta\dot{Q}(t)\geq0.
\end{equation}
Hence, for a time-independent $H$, a negative production rate for some $t$ is a rigorous indicator of the non-Markovian character of the dynamics. It is clear the presence of intervals with a strong negative production rate in Fig. 1.

\paragraph{Conclusions.---} 
We have presented a general thermodynamic framework for open quantum systems in contact with a thermal reservoir. This was done by identifying the nonequilibrium internal energy imposing suitable initial and asymptotic conditions, and the recovery of the standard weak-coupling result as an appropriate limit. The factorized initial condition was analyzed in detail and generalized to two natural extensions of correlated initial states. Furthermore, we have found that Markovian dynamics imply monotonically increasing entropy production.  This provides quantum non-Markovianity with a thermodynamic meaning, and allows for the introduction of new physical quantifiers of non-Markovianity.
Notably, all quantities in this approach can be inferred from measurements involving only system observables. At most, several preparations might be needed to determine $H_\SS^{\circledast}(t,\beta)$ and its derivatives, but no controlled reservoirs are required. This greatly simplifies the approach and opens the possibility to measure these strong coupling thermodynamic variables in the lab.

\begin{acknowledgments}

The author is grateful to M. Merkli and P. Strasberg for enlightening discussions. Financial support from Spanish MINECO grants FIS2017-91460-EXP, PGC2018-099169-B-I00, the CAM research consortium QUITEMAD S2018/TCS-4342, and US Army Research Office through grant W911NF-14-1-0103 is acknowledged.

\end{acknowledgments}

\onecolumngrid
\clearpage
\twocolumngrid

\section*{SUPLEMENTARY MATERIAL}

\appendix

\setcounter{figure}{0}
\setcounter{equation}{0}
\renewcommand*{\thefigure}{S\arabic{figure}}
\renewcommand*{\theequation}{S\arabic{equation}}

\subsection{I.\quad A qubit interacting with a composite spin-boson reservoir}
\label{app_A}
In this model we consider a qubit system with Hamiltonian given by $H_{\rm S}=\tfrac{\omega_0}{2}\sigma_{\SS}^z$. This qubit is coupled to some reservoir made up of another qubit (spin) with Hamiltonian $H_{\rm spin}$ and a continuum bosonic system with Hamiltonian $H_{\rm boson}=\int dk \omega(k) a^\dagger(k)a(k)$. The bosonic part of the reservoir is not directly coupled to the system, instead, the system interacts with the reservoir spin and this one interacts with the bosonic modes. We assume a dipolar-like coupling between system and reservoir (spin), 
\begin{equation}
V=\kappa (\sigma^+_{\SS}\sigma^-_{\rm R}+\sigma^-_{\rm S}\sigma^+_{\RR}),
\end{equation}
and the total reservoir Hamiltonian reads $H_{\RR}=H_{\rm spin}+H_{\rm boson}+V_{\rm spin-boson}$, with 
\begin{equation}
V_{\rm spin-boson}=\alpha \int dk g(k)\sigma^x_{\RR}[a(k)+a^\dagger(k)].
\end{equation}
In these equations, $a(k)$ denote bosonic anihilation operators, and $\sigma^{z,\pm}_{\SS}$, $\sigma^{z,x,\pm}_{\RR}$ stand for Pauli matrices for system and spin part of the reservoir, respectively. Thus, the system-reservoir interaction strength is mediated by $\kappa$, and the interaction between the reservoir spin and the bosonic continuum is assumed to be weak. In this circumstance, the exact (formal) convergence 
\begin{equation}
\rho_\SS\otimes \rho_{\RR,\beta}\xrightarrow{t\to\infty}\rho_{\SR,\beta}
\end{equation}
is rewritten approximately as
\begin{equation}
\rho_\SS\otimes \frac{e^{-\beta (H_{\rm spin}+H_{\rm boson})}}{Z_{\rm spin}Z_{\rm boson}}\xrightarrow{t\to\infty}\frac{e^{-\beta (H_\SS+H_{\rm spin})}\otimes e^{-\beta H_{\rm boson}}}{Z_{\rm S-spin}Z_{\rm boson}}
\end{equation}
with $Z_{\rm spin/boson}=\Tr\{\exp[-\beta H_{\rm spin/boson}]\}$, $Z_{\rm S-spin}=\Tr\{\exp[-\beta (H_\SS+H_{\rm spin})]\}$.

The weak reservoir spin-boson coupling $\alpha$ is assumed to be small enough such that the evolution of the system-spin density matrix $\rho_{\rm S-spin}$ is well approximated by the Davies semigroup. The Davies generator can be obtained by the standard procedure of second order expansion and secular approximation \cite{AlickiBook_SM,BrPe02_SM,Libro_SM}. To this end, we find the spectrum of $H_\SS+V+H_{\rm spin}$, which is $\{\omega_0,\kappa,-\kappa,-\omega_0\}$, where we have taken the resonance condition $\omega_0=\omega_1$ for simplicity. The spin-boson coupling is rewritten as 
\begin{widetext} 
\begin{equation}
V_{\rm spin-boson}=\alpha\int dk g(k)\sigma^x_{\RR}[a(k)+a^\dagger(k)]=\alpha\int dk g(k)\Big\{{\textstyle \sum_{\omega=|\omega_0\pm\kappa|}}[A(\omega)+A^\dagger(\omega)]\Big\}[a(k)+a^\dagger(k)],
\end{equation}
where $A(\omega)$ is the eigenoperator of $H_\SS+H_{\rm spin}+V$ with associated Bohr frequency $\omega$, $[H_\SS+H_{\rm spin}+V,A(\omega)]=-\omega A(\omega)$. Note that $A(-\omega)=A^\dagger(\omega)$. In coordinates,
\begin{equation}
A(\omega_0+\kappa)=\frac12\begin{pmatrix}
0 & 0 & 0 & 0\\
1 & 0 & 0 & 0\\
-1 & 0 & 0 & 0\\
0 & 1 & 1 & 0
\end{pmatrix}, \quad A(\omega_0-\kappa)=\frac12\begin{pmatrix}
0 & 0 & 0 & 0\\
1 & 0 & 0 & 0\\
1 & 0 & 0 & 0\\
0 & -1 & 1 & 0
\end{pmatrix}.
\end{equation}
Therefore, taking for the sake of definiteness $\omega_0\geq \kappa\geq0$, the Davies generator reads 
\begin{align}\label{mastereq}
\frac{d\rho_{\rm S-spin}(t)}{dt}=-i[H_\SS+H_{\rm spin},\rho_{\rm S-spin}(t)]+\sum_{\omega=\omega_0\pm\kappa} &\gamma(\omega)[\bar{n}(\omega)+1]\big[A(\omega)\rho_{\rm S-spin}(t)A^\dagger(\omega)-\tfrac12\{A^\dagger(\omega)A(\omega),\rho_{\rm S-spin}(t)\}\big]\nonumber \\
&\gamma(\omega)\bar{n}(\omega)\big[A^\dagger(\omega)\rho_{\rm S-spin}(t)A(\omega)-\tfrac12\{A(\omega)A^\dagger(\omega),\rho_{\rm S-spin}(t)\}\big].
\end{align}
Here, we have neglected the Lamb shift term, $\bar{n}(\omega)=[\exp(\beta \omega)-1]^{-1}$ is the mean number of bosons in the reservoir with frequency $\omega$, and $\gamma(\omega)=2\pi J(\omega)$. The spectral density of the bosonic continuum is given by  $J(\omega):=g^2[k(\omega)]\frac{dk(\omega)}{d\omega}$, with $k(\omega)$ the inverted function of the ``dispersion'' relation $\omega(k)$. 
\end{widetext}
Equation \eqref{mastereq}, which can be easily solved by matrix {exponentiation}, describes the relaxation 
\begin{equation}
\rho_{\rm S-spin}(0)=\rho_\SS\otimes\rho_{{\rm spin},\beta}\xrightarrow{t\to\infty}\frac{e^{-\beta (H_\SS+H_{\rm spin})}}{Z_{\rm S-spin}}
\end{equation}
and provides $\rho_{\rm S-spin}(t)$ for any $t$. The open system evolution is then given by taking partial trace on the reservoir spin
\begin{equation}
\rho_\SS(t)=\Tr_{\rm spin}[\rho_{\rm S-spin}(t)].
\end{equation}
These operations, despite tedious, can be done in an exact, analytical way, and allow for the computation of $H_\SS^\circledast(t)$ and so the internal energy and the thermodynamic entropy. The results are shown in the main text, Figure 1, for the qubit system initially in the ground state of $H_\SS$, with $\gamma(\omega_0\pm\kappa)=10^{-3}\omega_0,$ and $\kappa=0.9\omega_0$, as indicated there. 

\begin{figure}[h!]
	\includegraphics[width=0.7\columnwidth]{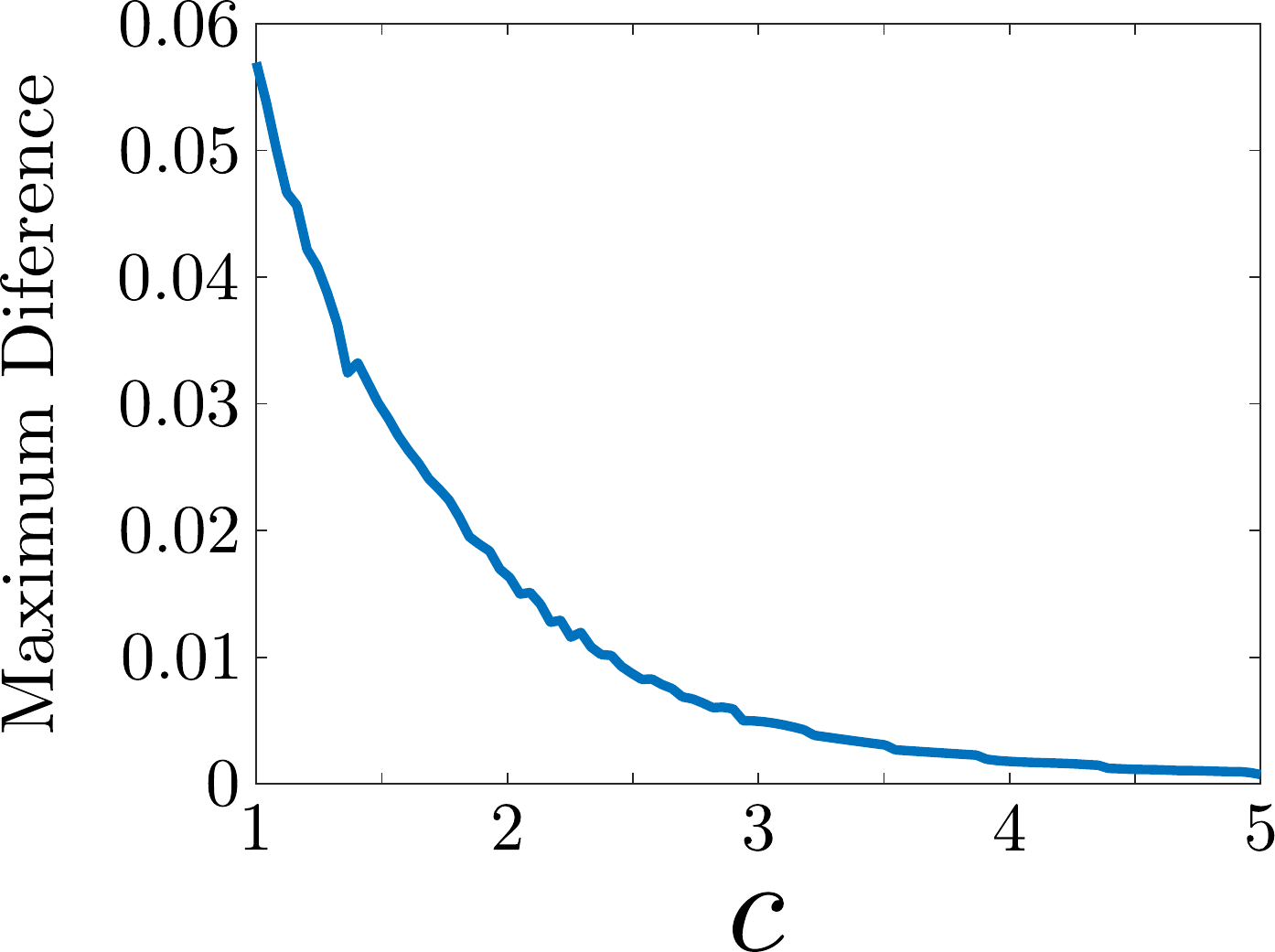}
	\caption{Maximum difference between internal energy and system Hamiltonian mean value, Eq. \eqref{dif}, as a function of the system-reservoir coupling strength, which is proportional to the inverse of $c$, Eq. \eqref{c}. The system is initially in the ground state of $H_{\SS}$. The expected convergence is shown.}
	\label{Fig:SM1}
\end{figure}

In addition, it is  worth to study the strong-weak coupling transition in the thermodynamic variables for this model. To this end, we can parametrize 
\begin{equation}\label{c}
\kappa=0.9/c, \text{ and } \gamma(\omega_0\pm\kappa)=10^{-3}\omega_0/c,
\end{equation}
so that we can approach the weak coupling condition (small $\kappa$) by increasing $c$, decreasing at the same time $\gamma(\omega_0\pm\kappa)$ in order not to jeopardize the Davies semigroup treatment. We plot in Fig.~\ref{Fig:SM1} the decay of the maximum difference between $E_{\rm U}$ and $\langle H_{\rm S}\rangle$,
\begin{equation}\label{dif}
\max_{t}|E_{\rm U}(t)-\langle H_{\rm S}(t)\rangle |,
\end{equation} 
as $c$ increases. We can see the expected convergence between both quantities.

\subsection{II.\quad Comparison with the nonequilibrium Hamiltonian of mean force approach}
\label{app_B}

\begin{figure}[t]
	\includegraphics[width=\columnwidth]{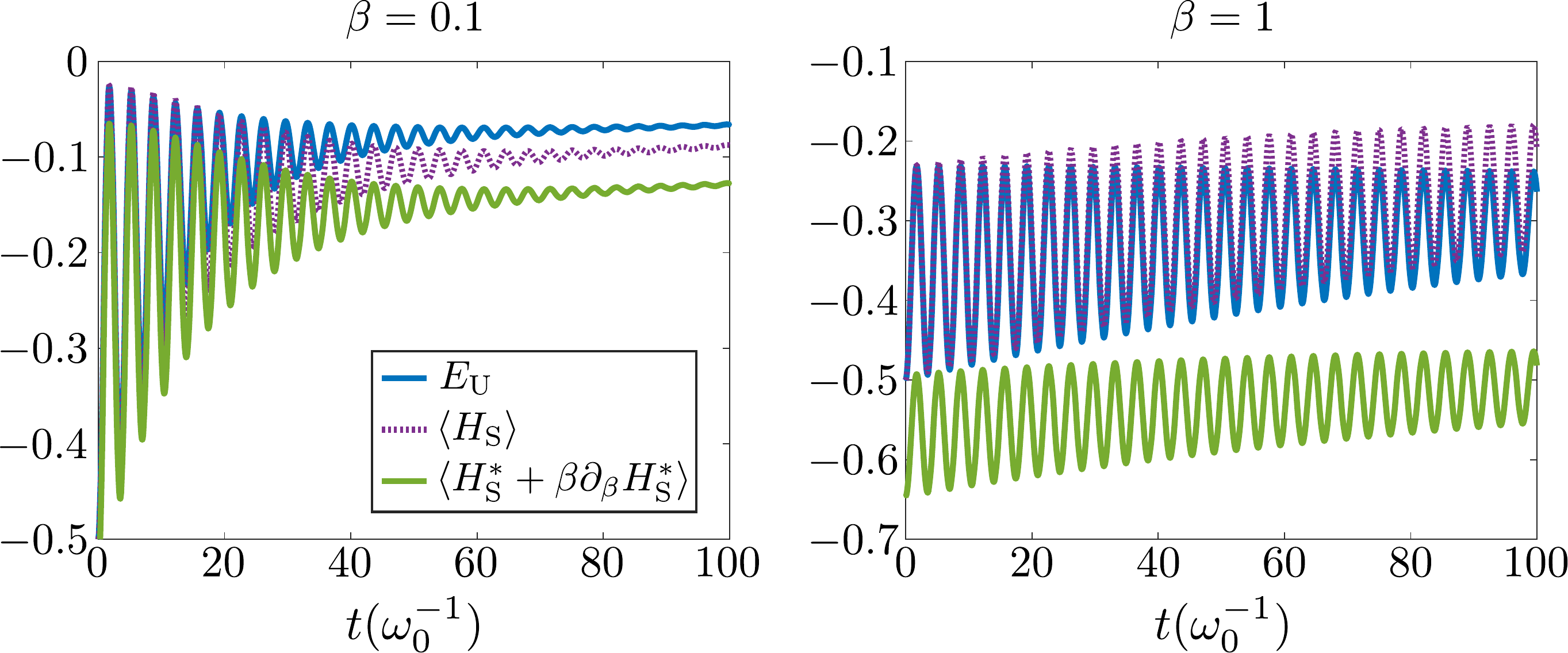}
	\caption{Internal energy $E_{\rm U}$ in comparison to $\langle H_{\SS}\rangle$ and $\langle H_{\SS}^*+\beta\partial_\beta H_\SS^*(\beta)\rangle$, Eq. \eqref{E**} for $\beta=0.1$ (left) and $\beta=1$ (right). The system is initially in the ground state of $H_{\SS}$.  The plots shows that $E_{\rm U}$ is generally closer to $\langle H_{\rm S}\rangle$ than $\langle H_\SS^*(\beta)+\beta \partial_\beta H_\SS^*(\beta) \rangle$, however the time-dependence displays a similar profile for the three values.}
	\label{Fig:SM2}
\end{figure}

\begin{figure}[t]
	\includegraphics[width=0.7\columnwidth]{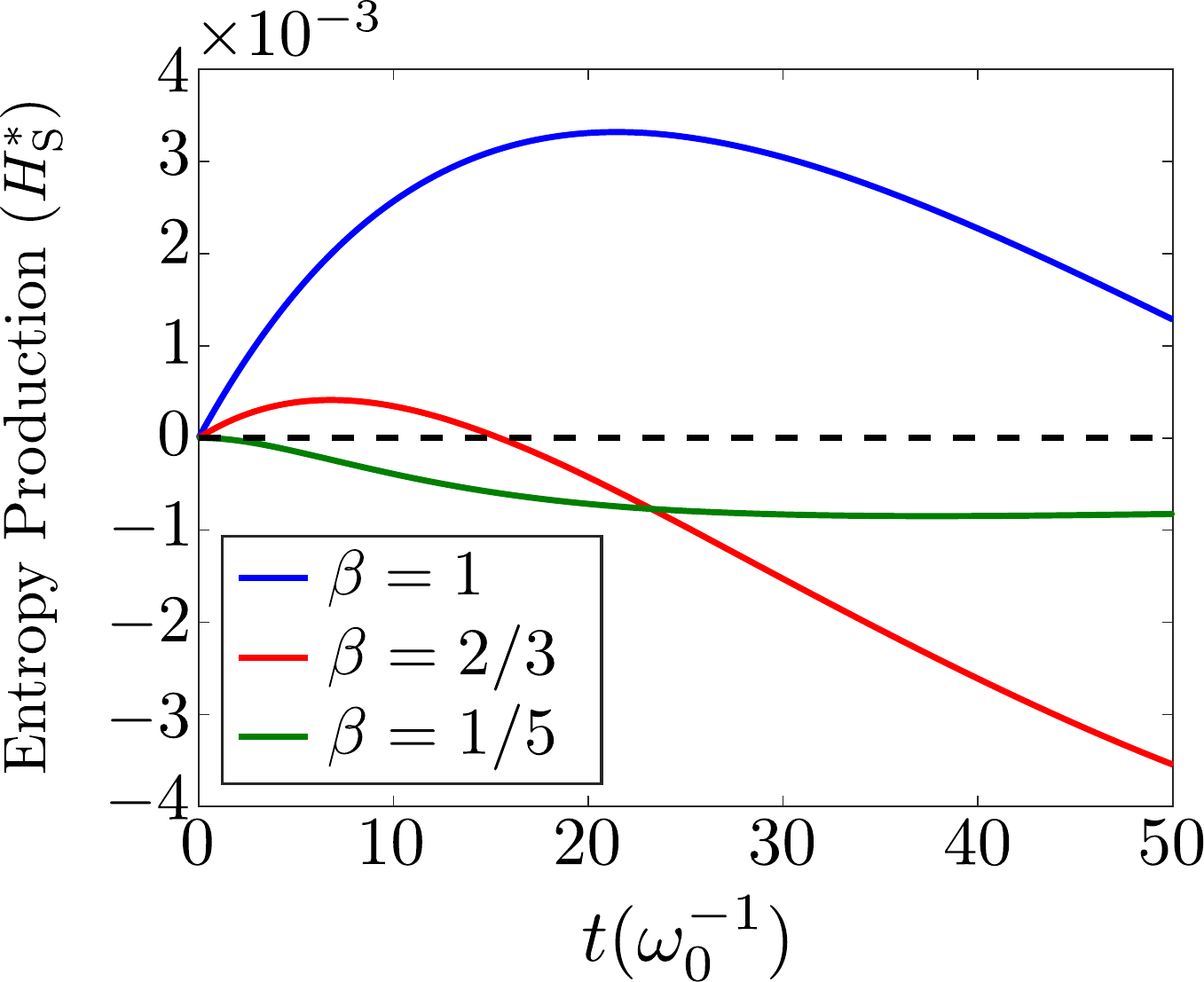}
	\caption{Entropy production $\Delta S^*(t)-\beta Q^*(t)$  in the approach of \cite{Seifert_SM,StrasbergPRE_SM,StragsberClass_SM}, see Eqs. \eqref{E**} and \eqref{S**}, for the initial state \eqref{prodthermal}, $\kappa=0.95\omega_0$, $\gamma(\omega_0\pm\kappa)=10^{-3}\omega_0$. Time intervals with negative values are clearly shown. The entropy production in the main text equals zero for this case $\Delta S(t)-\beta Q(t)=0$ (dotted black line).}
	\label{Fig:SM3}
\end{figure}

It is interesting to compare our results with the nonequilibrium approach based on the substitution $\rho_{\rm S, eq}\to \rho_{\rm S}(t)$ in the equilibrium equations (11)-(13) in terms of the Hamiltonian of mean force $H_{\SS}^*$ \cite{Seifert_SM,StrasbergPRE_SM,StragsberClass_SM}. For instance, for the internal energy, the equilibrium equation
\begin{equation}
 E_{\rm U}^*:=\Tr\{\rho_{\SS,\mathrm{eq}} [H_\SS^*(\beta)+\beta \partial_\beta H_\SS^*(\beta)]\}
\end{equation}
is generalized to
\begin{align}
E_{\rm U}^*(t):&=\Tr\{\rho_{\rm S}(t) [H_\SS^*(\beta)+\beta \partial_\beta H_\SS^*(\beta)]\}\\
&=\langle H_\SS^*(\beta)+\beta \partial_\beta H_\SS^*(\beta) \rangle.\label{E**}
\end{align}
In Fig.~\ref{Fig:SM2}, we present the computation of this internal energy, the internal energy $E_{\rm U}$, and the mean value of the system Hamiltonian $\langle H_{\rm S}\rangle$ for the qubit coupled to a composite spin-boson reservoir model. We may observe a similar temporal behavior for the three quantities, but $E_{\rm U}$ is generally closer to $\langle H_{\rm S}\rangle$ than $\langle H_\SS^*(\beta)+\beta \partial_\beta H_\SS^*(\beta) \rangle$. Notably, in the low temperature limit one has exactly the same result, as
%
\begin{align}
&\lim_{\beta\to\infty}H_{\rm S}^*(\beta)=\lim_{\beta\to\infty}H_{\rm S}^\circledast(t,\beta)=\begin{pmatrix}
\tfrac{\omega_0}{2}-\kappa & 0\\
0 & -\tfrac{\omega_0}{2}
\end{pmatrix},\\
&\lim_{\beta\to\infty}\partial_\beta H_{\rm S}^*(\beta)=\lim_{\beta\to\infty}\partial_\beta H_{\rm S}^\circledast(t,\beta)=\bm{0}.
\end{align}
Similarly, the thermodynamic entropy in the approach of \cite{Seifert_SM,StrasbergPRE_SM,StragsberClass_SM} reads
\begin{equation}
S^*(t):=\Tr\{\rho_{\SS}(t) [-\log\rho_{\SS}(t)+\beta^2\partial_\beta H_{\SS}^*(\beta)]\}. \label{S**}
\end{equation}
As commented in the main text, with these definitions, the entropy production $\Delta S^*(t)-\beta Q^*(t)$ has been shown to be positive for a restricted class of initial states \cite{StrasbergPRE_SM}. We can actually shown that this entropy production is negative in general. We consider again the qubit coupled to a composite spin-boson reservoir model, with an initial state given the product of individual canonical states, at the same temperature,
\begin{equation}\label{prodthermal}
\rho_\SS(0)=\rho_{\SS,\beta}\otimes\rho_{\RR,\beta}.
\end{equation}
For this initial condition the entropy production obtained in the main text gives a zero value. However, $\Delta S^*(t)-\beta Q^*(t)$ becomes negative for several temperatures, as shown in Fig.~\ref{Fig:SM3}. This suggests that the definitions \eqref{E**} and \eqref{S**} are a bad choice for internal energy and thermodynamic entropy under product initial conditions. Recently, a modification of this approach in \cite{Seifert_SM,StrasbergPRE_SM,StragsberClass_SM}, which does not suffer this problem, has been proposed by including the measurement apparatus in the formulation of the thermodynamic laws  \cite{StrasbergArxiv_SM}.

%
%


\begin{thebibliography}{00}
\bibitem{Footnote1} We write $E_{\rm U}$ instead the common notation $U$ to avoid any confusion with the unitary evolution operator.
\bibitem{Footnote3} 
Note that the change $H_{\rm S}\to H_{\rm S}^*$ in the definition \eqref{Hcir1} leads to the `symmetric' result $H_\SS^\circledast(0,\beta)=H_{\rm S}^{\circledast}(\infty,\beta)=H_{\rm S}^*$, and allows for a similar derivation of thermodynamic laws. However, it breaks the nice condition $E_{\rm U}(0)=\langle H_\SS(0)\rangle$ for the product state \eqref{InitialState} and requires the knowledge of $Z_\SR/Z_\RR$.
\bibitem{Footnote3bis}  In this approach, at equilibrium, we may write $Z_\SR=Z_{\SS}Z_{\RR}Z_{\rm int}$, with $Z_{\rm int}=Z_\SR/(Z_{\SS}Z_{\RR})$, so that $\mathcal{X}_\SR=\mathcal{X}_\SS+\mathcal{X}_\RR+\mathcal{X}_{\rm int}$ for $\mathcal{X}=E_U,F,S$, where $\mathcal{X}_{j}$ is the thermodynamic variable $\mathcal{X}$ computed from $Z_j$. Therefore, the strength of the coupling $V$ affects to $\mathcal{X}_{\rm SR}$ and $\mathcal{X}_{\rm int}$, but not to $\mathcal{X}_{\rm S}$ and $\mathcal{X}_{\rm R}$.
\bibitem{Footnote4} The CPTP map is formally given by the Kraus operators $K_{m,n}=\sum_{k}\sqrt{\lambda_{m|k}}\langle \psi_{\RR,m|k}| U(t,0)|\psi_{\RR,n|k}\rangle\Pi_k $, where $\rho_{\RR|k}=\sum_{m}\lambda_{m|k}|\psi_{\RR,m|k}\rangle\langle \psi_{\RR,m|k}|$ is the spectral decomposition of $\rho_{\RR|k}$ \cite{Cesar}.
\end{thebibliography}
\end{document}